\documentclass[fleqn,10pt]{wlscirep}

\usepackage[utf8]{inputenc}
\usepackage[T1]{fontenc}

\usepackage{graphicx}
\usepackage{xurl}
\usepackage{multirow}
\usepackage{amsmath,amssymb,amsfonts}
\usepackage{amsthm}
\usepackage[title]{appendix}
\usepackage{xcolor}
\usepackage{textcomp}
\usepackage{manyfoot}
\usepackage{booktabs}
\usepackage{algorithm}
\usepackage{algorithmicx}
\usepackage{algpseudocode}
\usepackage{listings}
\usepackage{array}
\usepackage{acro}
\usepackage{needspace}

\DeclareAcronym{emt}{
  short = EMT,
  long = electromagnetic transient
}
\DeclareAcronym{fd}{
  short = FD,
  long = fault detection
}

\DeclareAcronym{fc}{
  short = FC,
  long = fault classification
}

\DeclareAcronym{fl}{
  short = FL,
  long = fault localization
}

\DeclareAcronym{ed}{
  short = ED,
  long = event detection
}

\DeclareAcronym{ec}{
  short = EC,
  long = event classification
}

\DeclareAcronym{rf}{
  short = RF,
  long = random forest
}

\DeclareAcronym{mlp}{
  short = MLP,
  long = multilayer perceptron
}

\DeclareAcronym{gru}{
  short = GRU,
  long = gated recurrent unit
}

\DeclareAcronym{cnn}{
  short = CNN,
  long = convolutional neural network
}

\DeclareAcronym{resnet}{
  short = ResNet,
  long = residual network
}

\DeclareAcronym{mae}{
  short = MAE,
  long = mean absolute error
}

\theoremstyle{plain}

\theoremstyle{definition}

\theoremstyle{remark}

\title{EvEMTBench: An Open Benchmark for Machine Learning in Power System Protection}

\author[1,*]{Julian Oelhaf}
\author[2]{Georg Kordowich}
\author[3]{Christian Bergler}
\author[1]{Andreas Maier}
\author[2]{Johann J\"ager}
\author[1]{Siming Bayer}

\affil[1]{
Pattern Recognition Lab,
Friedrich-Alexander-Universit\"at Erlangen-N\"urnberg,
Erlangen 91058, Germany
}

\affil[2]{
Institute of Electrical Energy Systems,
Friedrich-Alexander-Universit\"at Erlangen-N\"urnberg,
Erlangen 91058, Germany
}

\affil[3]{
Department of Electrical Engineering, Media and Computer Science,
Ostbayerische Technische Hochschule Amberg-Weiden,
Amberg 92224, Germany
}

\affil[*]{julian.oelhaf@fau.de}

\begin{abstract}
Studies of machine-learning-based power system protection are difficult to compare because task definitions, measurement access, data partitions, metrics, and generalization conditions often differ. EvEMTBench addresses this gap with an open, executable, and versioned benchmark that fixes these evaluation choices while leaving model design open. Across four grids spanning 20--345\,kV, it defines 12 protection and event-analysis functions instantiated as 24 scored tasks and supports structured evaluation across observability conditions, predefined distribution shifts, and zero-shot and fine-tuned cross-grid transfer. Committed partitions, leakage controls, and reproducible reporting provide a common basis for comparing future methods. A reference evaluation spanning trivial, conventional, feature-based, and deep-learning baselines shows that wider observability is not uniformly beneficial, shifted conditions can reveal failures not apparent in-distribution, and cross-grid transfer is substantially stronger for fault detection than for fault localization. Protection-relevant diagnostics identify failure modes not apparent from primary metrics alone. EvEMTBench therefore makes generalization in machine-learning-based protection an explicit and reproducible evaluation problem.
\end{abstract}

\keywords{machine learning, power system protection, benchmark, reproducibility, electromagnetic transients}

\begin{document}

\flushbottom
\maketitle
\thispagestyle{empty}

\section*{Introduction}\label{sec:introduction}

Power system protection must make reliable and time-critical decisions across a growing range of operating and fault conditions. Inverter-based and distributed resources contribute to this variability by altering fault-current magnitudes and transient responses, including lower and electronically limited fault currents that depart from assumptions underlying established protection functions~\cite{cigre_joint_working_group_b5c626cired_protection_2015,hooshyar_microgrid_2017,haddadi_impact_2021,chowdhury_transmission_2021,quispe_transmission_2022}. Protection must nevertheless operate dependably by identifying and clearing faults when required, and securely by avoiding unnecessary operation during non-fault events such as switching or inrush phenomena~\cite{blackburn_protective_2014,abder_elandaloussi_practical_2023}. These requirements do not change when protection functions are implemented with machine learning. Evidence for machine-learning-based protection must therefore be grounded in clearly defined operating, sensing, and validation conditions rather than in test-set performance alone~\cite{mederer_verification_2025}.

Current evaluation practice makes such evidence difficult to interpret. A recent scoping review found substantial variation in grids, event types, sampling rates, observability assumptions, data splits, and evaluation metrics across protection studies~\cite{oelhaf_scoping_2025}. Shared datasets and released code remain uncommon, and robustness across sensing conditions and grid topologies is rarely assessed systematically. Controlled studies further show that data sparsity and degraded sensing can affect protection tasks differently~\cite{oelhaf_impact_2025,oelhaf_robustness_2026}. These evaluation choices are not secondary implementation details: topology, measurement access, temporal context, target construction, and split strategy determine the inference problem. Studies addressing the same nominal protection task may therefore evaluate materially different problems, so differences in reported performance cannot be attributed to model quality alone. Leakage and incomplete reporting can further inflate results and obstruct reproducibility~\cite{kapoor_leakage_2023,kapoor_reforms_2024}.

Even comparisons under matched conditions address only part of the problem. Learning-based protection methods are commonly developed within a single grid or simulated operating domain~\cite{oelhaf_scoping_2025}, and strong in-distribution performance does not establish that a model will retain its behavior on another network, voltage level, topology, operating condition, or sensing configuration. More generally, model rankings can change with the selected tasks, datasets, metrics, and ranking procedures~\cite{dehghani_benchmark_2021,maier-hein_why_2018}, while performance under distribution shift can differ substantially from standard in-distribution evaluation~\cite{koh_wilds_2021}. This distinction is particularly relevant to protection because fault signatures depend on the physical grid, its operating state, and the information available to the model. A benchmark for machine-learning-based protection must therefore support both comparison under matched conditions and evaluation beyond the distributions and grids used during development.

Shared data alone are therefore insufficient: the evaluation conditions must be defined consistently across methods. We introduce the \textbf{EvEMTBench Benchmark}, the executable and versioned evaluation component of the broader EvEMTBench project, built on the separately documented \textbf{EvEMTBench Dataset}~\cite{kordowich_simulation_2026}. The dataset provides \ac{emt} waveforms, labels, episode metadata, and topology information, while the benchmark specifies how machine-learning methods are evaluated on these data. The primary contribution is reusable evaluation infrastructure that fixes the comparison conditions while leaving model design open. Specifically, EvEMTBench contributes:

\begin{itemize}
    \item \textbf{A standardized executable specification} for 12 protection and event-analysis functions instantiated as 24 scored tasks, with fixed task definitions, valid-sample rules, observability views, metrics, and reporting requirements.
    \item \textbf{Structured generalization evaluation} spanning matched and predefined shifted conditions, three observability views, and zero-shot and fine-tuned cross-grid transfer.
    \item \textbf{Leakage-controlled and reproducible evaluation infrastructure} with committed group-aware partitions, training-only preprocessing, versioned artifacts, provenance records, and released evaluation software.
\end{itemize}

We demonstrate these capabilities through a common reference evaluation spanning trivial, conventional, feature-based, and deep-learning baselines. The evaluation addresses four research questions: \textbf{RQ1}, how reference performance varies across primary functions and grids under matched in-distribution evaluation; \textbf{RQ2}, how local, line, and global observability affect fault detection, classification, and localization; \textbf{RQ3}, how conclusions change between the in-distribution test partition and the predefined shifted \texttt{eval\_grid} condition on the same reference topology; and \textbf{RQ4}, which fault-analysis capabilities transfer across grids without target-grid training and how target-grid fine-tuning changes their performance. 

\section*{Background and Related Work}\label{sec:background}

This section reviews prior work on machine-learning-based power system protection, available datasets and benchmarks, and evaluation frameworks relevant to EvEMTBench.

\subsection*{Machine learning for power system protection}\label{sec:background_ml}

Learning-based protection involves more than recognizing faults. Models may need to distinguish normal operation, non-fault events, and faults before identifying fault type, affected element, or location, with controlled studies comparing model families for fault detection and line identification~\cite{oelhaf_systematic_2025} and for fault classification and localization~\cite{oelhaf_controlled_2026}. Proposed methods range from feature-based, wavelet-assisted, and zonal approaches~\cite{ekici_support_2012,livani_fault_2013,abdullah_ultrafast_2018,patnaik_modwt-xgboost_2021,poudel_zonal_2022} to hybrid deep-waveform and relay-embedded models~\cite{moradzadeh_hybrid_2022,jones_machine_2021} and topology-aware, spatiotemporal, or multi-task graph neural networks~\cite{chen_fault_2020,nguyen_spatial-temporal_2023,chanda_heterogeneous_2025,kordowich_graph_2025}. Reviews document applications across fault detection, classification, localization, asset protection, and emergency control~\cite{vaish_machine_2021,porawagamage_review_2024}. The literature therefore spans distinct protection objectives and substantially different model representations, motivating evaluation that is not tied to a single task or model family.

\subsection*{Datasets and benchmarks for power system protection}\label{sec:background_benchmarks}

A growing set of open resources supports machine-learning research in power systems and protection. Protection-relevant field data include curated event signatures, labeled substation oscillograms, and disturbance or fault recordings from IRTSD and RTE~\cite{wilson_grid_2024,evdakov_dataset_2026,ross_irtsd_2024,presvots_database_2024}, with RTE recordings already used for unsupervised event analysis and fault-inception detection~\cite{oelhaf_unsupervised_2025,oelhaf_fault_2026}. Complementary synthetic resources provide reference tasks, generated phasor data, graph-structured problems, and high-voltage electromagnetic-transient waveforms~\cite{zheng_multi-scale_2022,foggo_pmubage_2024,varbella_powergraph_2024,oelhaf_protect-90_2026}. Recent benchmarking efforts have also compared deep models on \ac{emt} streams under real-time latency constraints on a single system~\cite{abukhousa_latency-aware_2026}. The EvEMTBench Dataset provides controlled \ac{emt} waveforms, labels, metadata, and topology information across multiple grid families and event conditions~\cite{kordowich_simulation_2026}.

Relevant resources differ in signal representation, physical scope, task coverage, and the extent to which evaluation procedures are prescribed. Table~\ref{tab:related_benchmarks} compares EvEMTBench with PSML~\cite{zheng_multi-scale_2022}, pmuBAGE~\cite{foggo_pmubage_2024}, PowerGraph~\cite{varbella_powergraph_2024}, PROTECT-90~\cite{oelhaf_protect-90_2026}, and the latency-aware \ac{emt} benchmark of Abukhousa et al.~\cite{abukhousa_latency-aware_2026}.

\begin{table*}[ht]
\centering
\small
\caption{Comparison of selected public datasets and benchmarking resources relevant to EvEMTBench.}
\label{tab:related_benchmarks}

\begin{tabular}{@{}lcccccc@{}}
\toprule
\textbf{Property} & \textbf{PSML} & \textbf{pmuBAGE} & \textbf{PowerGraph} & \textbf{PROTECT-90} & \textbf{Latency-aware EMT} & \textbf{EvEMTBench} \\
\midrule
Public data & \(\checkmark\) & \(\checkmark\) & \(\checkmark\) & \(\checkmark\) & n.r. & \(\checkmark\) \\
Public evaluation code & \(\checkmark\) & -- & \(\checkmark\) & -- & n.r. & \(\checkmark\) \\
Point-on-wave \ac{emt} data & -- & -- & -- & \(\checkmark\) & \(\checkmark\) & \(\checkmark\) \\
Multiple grid systems & -- & n.r. & \(\checkmark\) & -- & -- & \(\checkmark\) \\
Prescribed evaluation protocol & \(\checkmark\) & -- & \(\checkmark\) & -- & -- & \(\checkmark\) \\
Cross-grid transfer & -- & -- & -- & -- & -- & \(\checkmark\) \\
\bottomrule
\end{tabular}

\vspace{2.5pt}
\parbox{0.97\linewidth}{\footnotesize
\textit{Note:} \(\checkmark\) indicates a verified property; ``--'' indicates a verified absence or a property outside the released resource's scope; ``n.r.'' indicates that the property was not reported or could not be verified.
}
\end{table*}

\subsection*{From evaluation frameworks to executable benchmarking}\label{sec:background_evaluation}

The differences summarized in Table~\ref{tab:related_benchmarks} concern not only the available data, but also how evaluation conditions are defined. Our evaluation framework identifies seven dimensions that must be made explicit for machine-learning-based protection results to be interpretable and comparable: the protection objective, physical system scope, observability and measurements, timing and decision windows, targets and valid samples, training and validation protocol, and evaluation outputs and diagnostics~\cite{oelhaf_standardized_2026}. The framework is intentionally independent of a particular dataset, grid, or model. However, specifying these dimensions does not by itself make evaluations directly comparable, because studies may still instantiate tasks, valid samples, measurement access, partitions, and metrics differently. Direct model comparison therefore requires these choices to be fixed in a shared, executable specification while model design remains open. The \textbf{EvEMTBench Benchmark} implements these requirements for a multi-grid \ac{emt} setting. To our knowledge, no existing protection-specific benchmark provides a public executable evaluation specification combining multiple fault- and event-analysis tasks, observability views, grid families, predefined shifted evaluation conditions, and cross-grid transfer on point-on-wave \ac{emt} data.

\section*{EvEMTBench Benchmark Design and Methods}\label{sec:methods}

This section defines the EvEMTBench Benchmark and the reference evaluation used in this study, including the dataset conditions, tasks, observability views, evaluation protocols, reference baselines, metrics, and reproducibility requirements.

\subsection*{Benchmark scope and terminology}\label{sec:methods_scope}

\textbf{EvEMTBench} denotes the overarching project. It comprises the \textbf{EvEMTBench Dataset}, which provides the \ac{emt} waveforms, labels, episode metadata, and topology information, and the \textbf{EvEMTBench Benchmark}, which specifies how machine-learning methods are evaluated on these data. The benchmark fixes the task definitions, valid-sample rules, observability views, data partitions, leakage controls, evaluation protocols, metrics, and reporting requirements while leaving model design open.

\begin{figure*}[t]
    \centering
    \includegraphics[width=\linewidth]{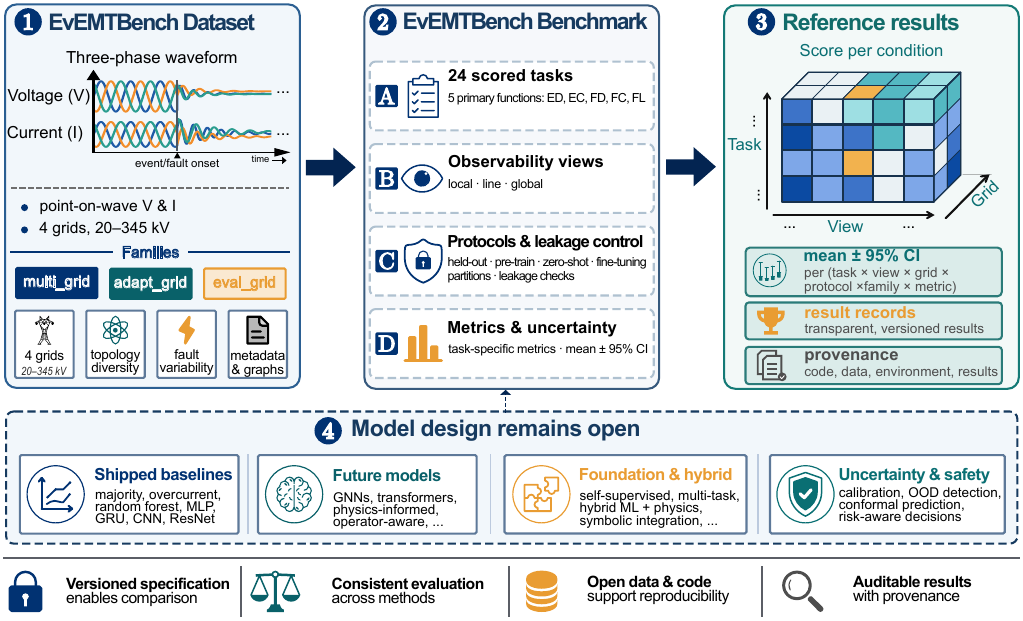}
    \caption{Overview of EvEMTBench. (1) The EvEMTBench Dataset provides point-on-wave voltage and current waveforms, topology information, and episode metadata for four grids spanning 20--345\,kV, organized into the \texttt{multi\_grid}, \texttt{adapt\_grid}, and \texttt{eval\_grid} families. (2) The EvEMTBench Benchmark fixes the scored tasks, observability views, evaluation protocols, metrics, group-aware partitions, and leakage controls. (3) Methods evaluated under this specification produce condition-specific scores with uncertainty estimates and machine-readable provenance records. (4) The benchmark is method-agnostic: the shipped baselines provide reproducible reference points, while the additional model families and evaluation directions shown are illustrative rather than results reported here.}
    \label{fig:benchmark_overview}
\end{figure*}

The experiments reported in this article constitute the \textbf{reference evaluation} of the benchmark. They apply a common suite of trivial, conventional, feature-based, and deep-learning baselines to establish reproducible reference results and illustrate the comparisons enabled by the benchmark. These results are not part of the benchmark definition. Figure~\ref{fig:benchmark_overview} summarizes the relationship between the dataset, benchmark specification, evaluated methods, and resulting evidence.

\subsection*{Dataset and generalization conditions}\label{sec:methods_dataset}

All experiments use the synthetic \ac{emt} dataset of Kordowich et al.~\cite{kordowich_simulation_2026}. The dataset paper documents the reference grids, component and event parameterization, PowerFactory simulations, raw record and topology formats, and technical validation. The four reference grids are established test systems from the literature: the CIGR\'E European medium-voltage benchmark grid~\cite{cigre_task_force_c60402_benchmark_2014}, a 110\,kV protection reference grid~\cite{lorz_interconnected_2023}, the IEEE 39-Bus New England system~\cite{athay_practical_1979}, and a simple double-line grid~\cite{ziegler_digitaler_2008}. Here we describe only the transformations that define the benchmark inputs and labels.

Each simulated episode (one retained simulation realization with its associated metadata) provides synchronized three-phase current and voltage measurements from the instrumented cubicles, sampled at \(9.6\,\mathrm{kHz}\) over the final \(0.5\,\mathrm{s}\) of a \(1.5\,\mathrm{s}\) simulation (4801 samples, both endpoints inclusive). The first second is discarded to exclude initialization transients. The benchmark uses \(50\,\mathrm{ms}\) windows of 480 samples with a \(5\,\mathrm{ms}\) step. With the default \(5\,\mathrm{ms}\) step, event inception is aligned with a window boundary. Window labels are derived deterministically from the episode metadata. For binary event-presence labels, a window is active if it overlaps the annotated event interval by any positive duration; windows entirely before event inception or after the annotated event interval are inactive. The overlap fraction is retained for diagnostic analysis.

The dataset paper denotes its three archive families as Multigrid, Adaptgrid, and Benchmark~\cite{kordowich_simulation_2026}. To avoid confusing the dataset family named Benchmark with the EvEMTBench Benchmark, this article uses the identifiers \texttt{multi\_grid}, \texttt{adapt\_grid}, and \texttt{eval\_grid}, respectively. Their generation and parameterization are documented in the dataset paper; here they define the data conditions summarized in Table~\ref{tab:dataset_families}. Their use within the evaluation protocols is specified below.

\begin{table*}[t]
\centering
\caption{Dataset families defining the EvEMTBench generalization conditions. The families differ in topology, operating state, and event-parameter variation.}
\label{tab:dataset_families}

\begin{tabular}{@{}
  >{\raggedright\arraybackslash}p{2.4cm}
  >{\raggedright\arraybackslash}p{6.1cm}
  >{\raggedright\arraybackslash}p{5.3cm}
@{}}
\toprule
\textbf{Family} &
\textbf{Variation} &
\textbf{Role} \\
\midrule

\texttt{multi\_grid}
& Topology, operating state, and event parameters vary
& Cross-grid pre-training and source-domain learning \\

\texttt{adapt\_grid}
& Topology is fixed; operating state and event parameters vary
& Target-grid training, validation, and in-distribution testing \\

\texttt{eval\_grid}
& Topology and operating state are fixed; event parameters follow predefined sweeps
& Evaluation-only shifted condition for held-out and transfer protocols \\

\bottomrule
\end{tabular}
\end{table*}

\subsection*{Benchmark tasks and valid-sample definitions}\label{sec:methods_tasks}

The windowed records defined above are mapped to 12 benchmark functions instantiated across their supported observability views, resulting in 24 function--view tasks. Five primary functions organize the main analysis. Three form the fault-analysis branch: \ac{fd}, \ac{fc}, and \ac{fl}; two form the event-analysis branch: \ac{ed} and \ac{ec}. Seven diagnostic functions cover fault grounding, faulted-phase identification, fault category, fault origin, event state, switch detection, and switch type. 

Not every window is valid for every function. Each task therefore has a frozen target derivation and valid-sample mask; windows outside that mask are excluded rather than converted into negative examples. For the fault-analysis branch, \ac{fd} is evaluated within fault episodes and predicts whether the fault is active in the current window. Fault-active windows are positive, whereas windows before fault inception or after the annotated fault interval are negative. Benign switching episodes are not used as negative examples for \ac{fd}. \Ac{fc} and \ac{fl} are evaluated only on fault-active windows: \ac{fc} predicts the fault class, while \ac{fl} predicts the fault position as a percentage of line length. Busbar faults are excluded from \ac{fl} because they do not have a unique position along a line.

The event-analysis branch uses a broader event definition. \Ac{ed} predicts whether any modeled fault or switching event is active, whereas \ac{ec} predicts the event class, including \texttt{no\_event} for inactive windows. The distinction between \ac{fd} and \ac{ed} is therefore one of evaluation scope: \ac{fd} evaluates fault activity within fault episodes, while \ac{ed} evaluates event activity across fault and switching episodes. The released reference-grid families contain no dedicated no-event episodes; negative samples for the event-analysis tasks therefore consist of inactive pre-event and post-event windows from fault and switching episodes.

Class vocabularies remain fixed across grids even when individual classes are unsupported. For \ac{fc}, the nine-class vocabulary contains seven shared fault classes and two high-impedance classes supported only on the medium-voltage grid. For \ac{ec}, the 23-class vocabulary includes grid-specific event types such as generator trips. Unsupported classes retain their indices but have zero support on the corresponding grid. Table~\ref{tab:task_spec} summarizes the target and valid-sample definitions of the five primary functions; definitions of the seven diagnostic functions are provided in the Supplementary Information.

\begin{table*}[t]
\centering

\caption{Target and valid-sample definitions for the five primary EvEMTBench functions. L, Li, and G denote local, line, and global observability, respectively.}
\label{tab:task_spec}

\begin{tabular}{@{}
  >{\raggedright\arraybackslash}p{2.7cm}
  >{\raggedright\arraybackslash}p{3.0cm}
  >{\raggedright\arraybackslash}p{8.3cm}
  >{\centering\arraybackslash}p{1.5cm}
@{}}
\toprule
\textbf{Function} &
\textbf{Target} &
\textbf{Valid evaluation windows} &
\textbf{Views} \\
\midrule

Fault detection (\acs{fd})
& Fault active / inactive
& All windows from fault episodes; fault-active windows are positive, while windows before inception or after the annotated fault interval are negative
& L, Li, G \\

Fault classification (\acs{fc})
& Fault class
& Fault-active windows with labels in the frozen nine-class fault vocabulary
& L, Li, G \\

Fault localization (\acs{fl})
& Position along line (\% of line length)
& Fault-active windows from line-fault episodes; busbar faults are excluded because no unique line position is defined
& L, Li, G \\

Event detection (\acs{ed})
& Event active / inactive
& All valid event-analysis windows; fault- or switching-active windows are positive, while inactive windows are negative
& G \\

Event classification (\acs{ec})
& Event class
& All valid event-analysis windows over the frozen 23-class vocabulary; inactive windows are labeled \texttt{no\_event}
& G \\

\bottomrule
\end{tabular}
\end{table*}

\subsection*{Observability and measurement views}\label{sec:methods_observability}

The benchmark varies measurement access through three nested views. For the fault-analysis tasks, local and line observability are centered on the ground-truth faulted element. For line faults, local observability contains the six voltage and current channels at one predefined terminal of the faulted line, while line observability adds the measurements from the opposite terminal to form a fixed 12-channel two-terminal view. Busbar faults use the documented adjacent-bay mapping. Global observability contains all main-line bay measurements and represents centralized or wide-area inference. Local and line views are therefore centered on the known faulted element and do not constitute fixed-relay selectivity tests.

Local and line input widths remain fixed across grids, whereas global width depends on topology. Fixed-input models can therefore transfer directly under local and line observability; global transfer requires a topology-aware representation. The reference models receive only the waveform channels defined by the selected observability view; explicit line parameters, operating-point variables, network-strength indicators, and topology metadata are not provided as model inputs.

\subsection*{Evaluation protocols, partitions, and leakage control}\label{sec:methods_protocols}

The benchmark defines four executable protocols: same-grid held-out evaluation (\texttt{held\_out}), source-corpus pre-training (\texttt{multi\_grid\_pretrain}), zero-shot cross-grid transfer (\texttt{transfer\_zeroshot}), and fine-tuned cross-grid transfer (\texttt{transfer\_finetune}). Training, model selection, and evaluation remain separated throughout, and the \texttt{eval\_grid} family is used only for final target-grid evaluation.

Under held-out evaluation, models are developed and evaluated on the same reference topology. For each grid, the \texttt{adapt\_grid} family is divided into training, validation, and in-distribution test partitions using a \(70/15/15\%\) split. Partitioning is performed at episode level using \texttt{sample\_id} and is stratified by event type, so all windows derived from one \ac{emt} episode remain in the same partition. Models are fitted on the training partition, while the validation partition is used for early stopping and model selection. The same committed partitions are used across all five experimental seeds.

Two evaluation conditions are reported for each held-out model. The \texttt{adapt\_grid} test partition measures in-distribution generalization under the same data family used for model development. The complete \texttt{eval\_grid} family provides a predefined shifted condition on the same reference topology. Its operating conditions and event-parameter sweeps differ from those of \texttt{adapt\_grid}; depending on the task, supported classes, localization positions, or negative-class composition may also differ. The comparison therefore reflects the combined shift defined by this evaluation family and should not be interpreted as a single-factor test. Neither evaluation condition is used to estimate preprocessing parameters, select thresholds, perform early stopping, tune hyperparameters, or select models.

\paragraph{Cross-grid transfer.}

Cross-grid transfer uses \texttt{multi\_grid} source corpora at 20, 110, and 345\,kV. These corpora contain randomized topologies that are distinct from the four target reference topologies; no target reference topology appears in source-corpus training, validation, or test partitions. Each source corpus is divided into training, validation, and test subsets using a \(70/15/15\%\) split grouped by generated topology. All simulations of one topology, and all line episodes derived from the same simulation, remain in the same subset. The committed partitions include machine-readable audits verifying that no topology, simulation, or episode occurs in more than one subset.

In zero-shot transfer, a deep baseline is trained on the source-corpus training partition and selected using only the corresponding source validation partition. The selected model is then applied directly to the target grid's \texttt{eval\_grid} family without using target-grid training or validation data. In fine-tuned transfer, the source-pre-trained checkpoint is adapted using the target grid's \texttt{adapt\_grid} training partition, and model selection uses the corresponding target validation partition before evaluation on the target grid's \texttt{eval\_grid} family. Fine-tuning updates all model parameters under the same optimization regime used for the from-scratch reference models. Because the standard \texttt{held\_out} and transfer protocols use different input normalization, their direct comparison does not isolate the effect of pre-training. For the control comparison reported in RQ4, target-only models are trained from random initialization using the same per-window normalization, deep-baseline candidate pool, committed target-grid splits, optimization settings, seeds, and validation-based selection as the fine-tuned models. This control leaves the benchmark protocols and version identifiers unchanged.

Transfer is restricted to deep baselines with transferable model weights and to local and line observability, whose input widths remain constant across grids. These protocols evaluate transfer across grids, voltage levels, and operating conditions with a fixed input width; they do not test fully topology-agnostic models.

\paragraph{Leakage prevention.}

Leakage controls are enforced at the partitioning, preprocessing, and model-selection levels. Episodes rather than windows are the unit of partitioning for the reference-grid data, preserving the dependence structure among windows from the same episode~\cite{roberts_cross-validation_2017}; source-corpus partitions additionally group by generated topology and simulation as described above. Dataset-level preprocessing parameters, including channel-standardization statistics, class weights, regression-target scaling, and overcurrent thresholds, are estimated from the relevant training partition only. Per-window normalization used by the pre-training and transfer protocols depends only on the window being transformed and does not estimate parameters across evaluation samples. The evaluation software checks partition disjointness before scoring.

\subsection*{Reference baselines and model selection}\label{sec:methods_models}

The reference suite spans trivial, conventional, feature-based, and deep-learning approaches, as summarized in Table~\ref{tab:baselines}. It is intended to provide reproducible reference points rather than an exhaustive comparison or a performance ceiling. Model-specific hyperparameters are fixed in advance; no task-specific hyperparameter search is performed.

\begin{table}[t]
\centering
\caption{Reference baseline suite. Complete configurations are reported in the Supplementary Information.}
\label{tab:baselines}
\begin{tabular}{@{}lll@{}}
\toprule
\textbf{Baseline} &
\textbf{Type} &
\textbf{Input representation} \\
\midrule
Majority
& Trivial
& Training labels or regression targets \\

Threshold
& Conventional
& Local phase currents \\

\Ac{rf}
& Feature-based
& Per-channel waveform statistics \\

\Ac{mlp}
& Deep
& Flattened waveform window \\

\Ac{gru}
& Deep
& Waveform sequence \\

\Ac{cnn}
& Deep
& Waveform sequence \\

\Ac{resnet}
& Deep
& Waveform sequence \\
\bottomrule
\end{tabular}
\end{table}

The trivial baseline predicts the most frequent training class for classification tasks and the mean training target for \ac{fl}. It provides a task-specific reference value. The conventional baseline is an overcurrent threshold defined only for binary local \ac{fd}. It computes the maximum root-mean-square current across the three phases and applies a scalar threshold determined from the training partition by maximizing balanced accuracy over 200 quantiles of the observed training statistic.

The \ac{rf} provides a classical feature-based reference. It contains 200 trees, uses balanced class weights for classification, and operates on seven summary statistics computed independently for each channel: mean, standard deviation, minimum, maximum, peak-to-peak range, root-mean-square value, and maximum absolute value. The four deep baselines operate directly on waveform windows. The \ac{mlp} uses a flattened input, the \ac{gru} processes the waveform sequentially, and the \ac{cnn} and \ac{resnet} use one-dimensional convolutions. 

For classification, the deep models minimize cross-entropy with inverse-frequency class weights derived from the training partition. Under the \texttt{held\_out} protocol, waveform inputs are standardized only using channel statistics estimated from the training partition. The \texttt{multi\_grid\_pretrain} and transfer protocols first apply a per-window, per-channel z-score over the time axis using statistics from that window alone, followed by the same training-set-based channel standardization. For \ac{fl}, the line-position target is standardized using the mean and standard deviation of the training targets and transformed back to its original percentage-of-line scale before evaluation.

\paragraph{Training and selection.}

The four deep baselines use the same optimization settings: AdamW with an initial learning rate of \(10^{-3}\), weight decay of \(10^{-4}\), and minibatches of 256 windows. Optimization proceeds for at most 60 epochs with cosine annealing using \(T_{\max}=60\). Early stopping monitors validation loss with a patience of 12 epochs, and the checkpoint with the lowest validation loss is restored.

Each learned baseline is trained with five seeds. Model capacity is not equalized across architectures. Hardware information and aggregate compute requirements are reported in the Supplementary Information.

Headline model selection does not involve further hyperparameter tuning. Within each deep baseline, early stopping selects the checkpoint using validation loss. Where the main text reports one learned reference per evaluation cell, the candidate learned baseline with the best corresponding validation metric is then selected as the headline reference. The trivial and conventional baselines do not participate in this selection. The selected model is subsequently evaluated under the corresponding evaluation condition; neither the in-distribution test partition nor the \texttt{eval\_grid} family influences selection. Full results for all candidate baselines are retained in the Supplementary Information.

\subsection*{Evaluation metrics and uncertainty}\label{sec:methods_evaluation}

Binary and multiclass tasks use balanced accuracy as the headline metric, with macro-\(\mathrm{F}_1\) and accuracy reported as secondary metrics. Binary detection tasks additionally report task-specific miss and false-positive rates. For \ac{fd}, the missed-detection rate, \(\mathrm{MDR} = \mathrm{FN}/(\mathrm{FN}+\mathrm{TP})\), and false-alarm rate, \(\mathrm{FAR} = \mathrm{FP}/(\mathrm{FP}+\mathrm{TN})\), quantify missed fault-active windows and false-positive pre- or post-fault windows within fault episodes; they do not assess false trips during benign switching or other non-fault disturbances. For \ac{ed}, the corresponding rates quantify missed events and false event alarms. Other binary diagnostic tasks are reported using balanced accuracy, macro-\(\mathrm{F}_1\), and accuracy without these detection-specific rates.

For multiclass tasks, balanced accuracy is the mean recall over classes supported in the evaluated set. The frozen macro-\(\mathrm{F}_1\) metric is averaged over the full class vocabulary, with unsupported classes contributing zero; a companion macro-\(\mathrm{F}_1\) over supported classes averages only over classes present in the evaluated set. Per-class \(\mathrm{F}_1\) is reported as undefined for classes with zero support, whereas supported classes without predicted positives receive an \(\mathrm{F}_1\) score of zero. For \(K\) supported classes, the corresponding analytical balanced-accuracy reference is \(1/K\).

Fault localization reports \ac{mae} as a percentage of line length, with root-mean-square error, median absolute error, and \(R^2\) as secondary metrics. Localization error is also stratified by true line position. For each detected fault episode, the \ac{fd} latency analysis identifies the first fault-active window with a correct positive prediction and records its start-time offset from fault inception. Because the complete \(50\,\mathrm{ms}\) window must be observed before inference, the corresponding decision latency is this offset plus \(50\,\mathrm{ms}\). Latency is conditional on successful detection and is reported alongside the episode-level detection rate.

\paragraph{Uncertainty and comparisons.}

Stochastic results are reported as means across five matched seeds with two-sided 95\% confidence intervals based on Student's \(t\) distribution with four degrees of freedom. Because all seeds use the same committed partition, these intervals quantify training stochasticity rather than uncertainty from resampling episodes, operating conditions, or grids. Per-seed values are retained in the result artifacts.

Given the five-seed design, comparisons are interpreted descriptively rather than as formal hypothesis tests. The intervals characterize variation across training runs but do not establish equivalence or statistical superiority between methods. No confirmatory significance claims are made across the benchmark matrix.

For classification tasks, learned results below the corresponding majority-baseline value are retained and flagged rather than discarded. For localization, the training-mean predictor is a diagnostic reference rather than a pass--fail threshold.

\subsection*{Benchmark versioning, reporting, and reproducibility}\label{sec:methods_reproducibility}

Benchmark versions identify the artifacts that determine score comparability. Changes to task labels, valid-sample masks, class vocabularies, observability definitions, windowing, metrics, or committed partitions require a version change. Additive extensions such as new grids, baselines, diagnostics, or optional protocols may leave existing evaluation cells unchanged. If a revision affects only specific protocols, unchanged artifacts and results retain their existing version identifiers.

Results are indexed by task, observability view, grid, protocol, evaluation family, and metric; no aggregate score is defined. Each run records the benchmark version, code commit, software dependencies, model configuration, parameter count, random seed, and leakage-control status; the committed split audits record the corresponding partition hashes. The current benchmark release is EvEMTBench \texttt{v1.1.0}. It retains the unchanged \texttt{held\_out} splits and reference results from \texttt{v1.0.0}, while the topology-grouped \texttt{multi\_grid\_pretrain} splits and dependent pre-training and transfer results are versioned \texttt{v1.1.0}. The release contains the corresponding committed splits, preprocessing pipeline, baseline implementations, evaluation software, result schema, leakage checks, software environment, and reference results.

\section*{Results}\label{sec:results}

\begin{figure*}[t]
    \centering
    \includegraphics[width=\linewidth]{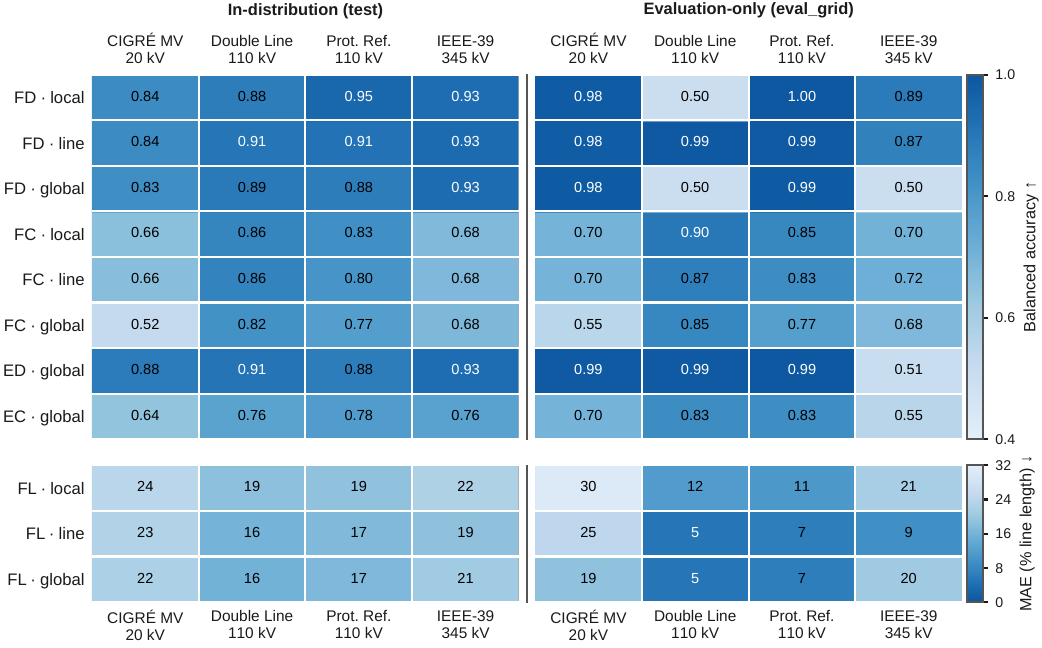}
    \caption{Main result synthesis for the \texttt{held\_out} protocol. Each cell shows the five-seed mean for the learned baseline selected using validation results for that function, observability view, and grid, evaluated on both the in-distribution test partition and the evaluation-only \texttt{eval\_grid} family. Classification and detection use balanced accuracy; fault localization uses mean absolute error as a percentage of line length. \Ac{ed} and \ac{ec} are evaluated only under global observability. Corresponding 95\% confidence intervals and complete per-baseline results are provided in the Supplementary Information. The \texttt{held\_out} results retain their \texttt{v1.0.0} identifiers in the current EvEMTBench release.}
    \label{fig:synthesis}
\end{figure*}

\begin{figure*}[t]
    \centering
    \includegraphics[width=0.94\linewidth]{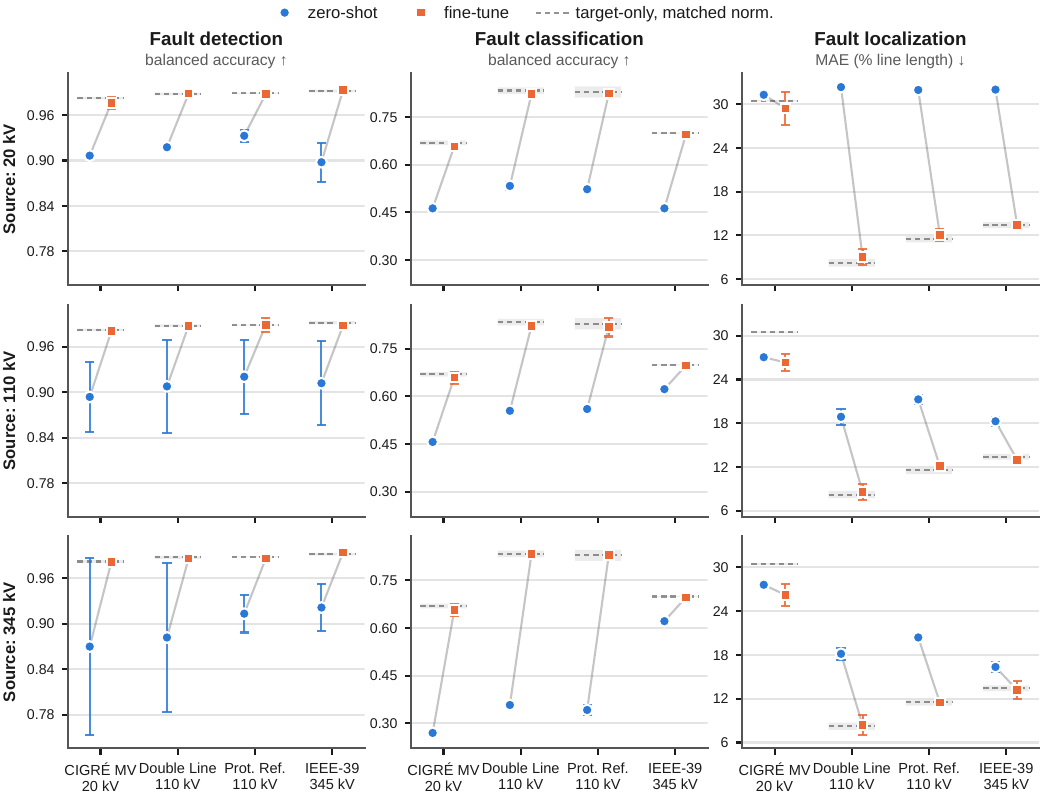}
    \caption{Cross-grid transfer under line observability for EvEMTBench \texttt{v1.1.0}. Rows denote the 20, 110, and 345\,kV pre-training sources, and columns the transferable fault-analysis functions. Paired markers compare zero-shot and fine-tuned transfer; dashed lines and shaded bands show the source-independent target-only reference under matched per-window normalization and deep-baseline selection. Zero-shot models are selected on source validation data, while fine-tuned and target-only models use target-grid validation data; \texttt{eval\_grid} is used only for evaluation. Markers and bands show five-seed means and 95\% confidence intervals; complete results are provided in the Supplementary Information.}
    \label{fig:transfer_overview}
\end{figure*}

The reference evaluation addresses RQ1--RQ4 in sequence, followed by protection-relevant diagnostics. Figure~\ref{fig:synthesis} summarizes held-out performance for the five primary functions, while Fig.~\ref{fig:transfer_overview} presents cross-grid transfer. Unless stated otherwise, stochastic results are reported as means over five seeds with 95\,\% confidence intervals; complete per-baseline results and model-selection records are provided in the Supplementary Information.

\subsection*{RQ1: Reference performance varies across functions and grids}
\label{sec:results_difficulty}

Reference performance differs substantially across the primary functions and grids. Most validation-selected learned references outperform the corresponding trivial baseline, but the improvement varies across tasks and grids. Detection is comparatively strong, classification is more grid-dependent, and fault localization remains the most difficult of the five primary functions for the evaluated baseline suite.

Among the fault-analysis functions, \ac{fd} achieves the strongest in-distribution performance. Across grids and observability views, the selected learned references reach balanced accuracy of approximately \(0.83\)--\(0.95\). \Ac{ed}, evaluated under global observability, shows similarly high reference performance, with balanced accuracy of approximately \(0.88\)--\(0.93\).

Classification is more variable across grids. For \ac{fc}, balanced accuracy reaches approximately \(0.86\) on the Double Line grid but only \(0.52\)--\(0.67\) on the CIGR\'E medium-voltage grid. The CIGR\'E medium-voltage grid also supports a larger fault-class vocabulary, including two high-impedance classes absent from the high- and extra-high-voltage grids. Because balanced accuracy averages recall over the classes supported on each grid, the analytical reference level differs accordingly: seven \ac{fc} classes are supported on the high- and extra-high-voltage grids and nine on the medium-voltage grid. A similar detection--classification difference appears for the event-analysis functions. Whereas \ac{ed} remains comparatively strong, \ac{ec} reaches balanced accuracy of approximately \(0.64\)--\(0.78\) across the four grids.

Fault localization has the weakest reference performance. The validation-selected learned references yield in-distribution \ac{mae} of approximately \(16\)--\(24\%\) of line length. The most difficult condition occurs on the CIGR\'E medium-voltage grid under local observability, where the error is approximately \(23.9\%\) and remains close to the corresponding training-mean reference. The selected learned reference therefore provides little improvement over the training-mean predictor in this condition.

\subsection*{RQ2: Observability affects protection functions differently}\label{sec:results_observability}

Increasing measurement access from local to line or global observability does not consistently improve reference performance. The effect differs by protection function, and independently selected headline results may correspond to different learned baselines across observability views. Model-matched comparisons are provided in the Supplementary Information.

For \ac{fd}, performance is generally similar across the three observability views, with no consistent improvement as measurement access increases. On the Protection Reference grid, balanced accuracy is \(0.949\) under local observability and \(0.878\) under global observability, although the two conditions select different baselines. \Ac{fc} likewise shows no consistent improvement with wider observability. On the CIGR\'E medium-voltage grid, balanced accuracy decreases from \(0.656\) under local observability to \(0.520\) under global observability. On the Double Line grid, the corresponding values are \(0.855\) and \(0.818\). The global estimate on the Double Line grid also shows comparatively large seed-to-seed variation, with a 95\% confidence-interval half-width of approximately \(0.087\), so the difference between the point estimates should be interpreted cautiously.

Fault localization shows a different pattern. Wider observability generally reduces localization error on the high-voltage grids. On the Double Line grid, \ac{mae} decreases from \(18.6\%\) under local observability to \(16.1\%\) under line observability and \(15.9\%\) under global observability, with similar trends on the other high-voltage grids. Across the reference evaluation, observability therefore has no monotonic effect on \ac{fd} or \ac{fc}, whereas \ac{fl} generally improves when measurements from both line terminals or the wider grid are available.

\subsection*{RQ3: Evaluation-family shifts are task dependent}\label{sec:results_shift}

Performance under the evaluation-only \texttt{eval\_grid} family is not uniformly lower than on the in-distribution test partition. For fault localization on the high-voltage grids, the shifted condition often yields lower error. On the Double Line grid under line observability, \ac{mae} decreases from \(16.1\%\) to \(5.2\%\), well below the corresponding training-mean reference of approximately \(31\%\). The \texttt{eval\_grid} family in this case uses a fixed operating point and a discrete five-position localization sweep, so the target distribution differs from that of the in-distribution test partition. The lower error should therefore not be interpreted as evidence of improved robustness under distribution shift. On the CIGR\'E medium-voltage grid, local localization remains close to the training-mean reference in both families, increasing from \(23.9\%\) to \(30.5\%\), whereas under global observability the error decreases from \(22.2\%\) to \(18.9\%\).

Detection shows a different response to the shift in several cells. Where validation selects a deep waveform baseline, performance generally remains high under the shift, including balanced accuracy of approximately \(0.99\) on the Double Line grid under line observability and across the Protection Reference grid. The \ac{rf}, however, is selected in eight of the twelve \ac{fd} cells and falls to approximately \(0.50\) balanced accuracy in several corresponding \texttt{eval\_grid} evaluations, including the local and global views on the Double Line grid and the global view on IEEE-39. The global \ac{ed} reference on IEEE-39 similarly falls to approximately \(0.51\). Other selected \ac{rf} cells retain high balanced accuracy, including approximately \(0.98\) on the CIGR\'E medium-voltage grid and \(0.87\)--\(0.89\) under local and line observability on IEEE-39. 

The conventional overcurrent threshold provides an additional reference for shifted local \ac{fd}. It reaches balanced accuracy of \(0.916\) on the Double Line grid compared with \(0.499\) for the selected learned reference, and \(0.914\) on IEEE-39 compared with \(0.886\). These cases do not establish general superiority of the conventional method, but they show that the effect of the predefined evaluation-family shift depends on the task and selected baseline.

\subsection*{RQ4: Cross-grid transfer differs by protection function}\label{sec:results_transfer}

Cross-grid transfer varies substantially across the three fault-analysis functions. \Ac{fd} transfers most consistently: outside the large-voltage-gap failures described below, zero-shot balanced accuracy is approximately \(0.87\)--\(0.98\), while fine-tuned performance reaches approximately \(0.95\)--\(0.99\). \Ac{fc} is less stable across source--target pairs but generally improves with target-grid fine-tuning. Under line observability, zero-shot models pre-trained at 345\,kV reach only \(0.27\)--\(0.36\) balanced accuracy on the three lower-voltage targets, increasing to \(0.66\)--\(0.83\) after fine-tuning.

Fault localization transfers least effectively. Under local observability, zero-shot \ac{mae} remains approximately \(27\)--\(34\%\) of line length across target grids and pre-training sources, close to the corresponding training-mean references of approximately \(30\)--\(31\%\). Fine-tuning reduces line-view error to approximately \(8\)--\(14\%\) on the high- and extra-high-voltage grids, whereas the medium-voltage grid remains more difficult, with line-view error of approximately \(26\%\).

To assess final target-grid performance, we compare fine-tuned models with target-only models under matched per-window normalization and deep-baseline selection. Across 72 source--target comparisons spanning \ac{fd}, \ac{fc}, and \ac{fl} under both transferable views, the 95\% confidence intervals overlap in 66 cells; of the remaining six, four favor fine-tuning and two favor target-only training. For line-view localization, matched target-only \ac{mae} is \(8.2\%\), \(11.5\%\), and \(13.4\%\) on the Double Line, Protection Reference, and IEEE-39 grids, respectively, closely matching the fine-tuned results. Cross-grid pre-training therefore provides zero-shot capability but no consistent improvement in final target-grid performance once target-grid data are available. Complete results are provided in the Supplementary Information.

Transfer also depends on the source grid and observability view. Under local observability, zero-shot \ac{fd} falls to approximately \(0.50\) balanced accuracy for the 345\,kV source on all lower-voltage targets and for the 20\,kV source on IEEE-39, whereas line-view transfer is more stable. For zero-shot localization, the 345\,kV source yields the lowest error on the high- and extra-high-voltage targets, while the 110\,kV source performs best on the medium-voltage target. 

\subsection*{Protection-relevant diagnostics identify failure modes}\label{sec:results_diagnostics}

The diagnostics provide additional context for several headline results. Although \ac{fd} shows the strongest reference performance among the primary fault-analysis functions, the selected references still miss approximately \(19\)--\(26\%\) of in-distribution fault-active windows on the CIGR\'E medium-voltage and Double Line grids. Balanced accuracy alone therefore does not capture the frequency of missed positive decisions within fault episodes.

The near-chance shifted-family \ac{fd} results correspond to a specific failure mode rather than random behavior. In these cells, the selected models predominantly predict the positive class, yielding missed-detection rates close to zero and false-alarm rates close to one on the pre- and post-fault windows included as negative samples. A balanced accuracy near \(0.50\) therefore reflects a loss of discrimination within the \ac{fd} evaluation set. Because benign switching and other non-fault disturbances are excluded from this task, these false-alarm rates do not measure protection security against such events.

Uncertainty also varies substantially across evaluation conditions. Several global-view cells on IEEE-39 show wide seed-to-seed intervals, whereas others are comparatively stable. Where intervals are wide, the reference results do not support a clear ranking between candidate baselines. Complete confidence intervals, per-seed results, and model-selection records are provided in the Supplementary Information; consistent with the five-seed design, these comparisons are interpreted descriptively rather than as evidence of statistical superiority.

\section*{Discussion}\label{sec:discussion}

The reference evaluation shows that conclusions about model performance depend on the protection function, measurement access, evaluation family, and generalization protocol. These dependencies determine how protection results and generalization claims should be interpreted and which results remain directly comparable.

\subsection*{Implications for power system protection}

Machine-learning-based protection performance cannot be characterized by a single score or model ranking. Strong fault-detection performance does not imply equally strong fault classification or localization, and the relative difficulty of these functions changes across grids. Evaluation should therefore preserve the protection function and its associated error modes rather than aggregate performance across tasks. This distinction is particularly important for protection, where missed operation, unnecessary operation, incorrect fault identification, and localization error have different operational consequences.

Observability is likewise part of the inference problem rather than a secondary implementation choice. Wider measurement access does not consistently improve \ac{fd} or \ac{fc} for the evaluated baselines, whereas \ac{fl} generally benefits from measurements at both line terminals or across the grid. Comparisons between methods are therefore meaningful only when measurement access is matched. At the same time, the local and line views used here are centered on the known faulted element. Their results characterize inference given measurements associated with that element and should not be interpreted as fixed-relay selectivity or out-of-zone fault tests.

The diagnostic results further show why headline metrics require protection-specific interpretation. Several shifted \ac{fd} failures correspond to predominantly positive predictions rather than random decisions, while a conventional overcurrent threshold outperforms the selected learned reference in some shifted conditions. This does not establish general superiority of conventional protection, but supports retaining established protection references alongside learned models. Moreover, the \ac{fd} false-alarm rate is computed from inactive windows within fault episodes; because benign switching and other non-fault disturbances are excluded, these results do not establish comprehensive protection security. Claims about dependable and secure operation therefore require evidence beyond aggregate performance metrics.

\subsection*{Implications for generalization and transfer}

The evaluation-family shift and cross-grid transfer evaluate distinct forms of generalization. Within a fixed reference topology, the \texttt{eval\_grid} family tests behavior under a predefined multidimensional shift in operating and event conditions. Its results show that a withheld distribution is not inherently more difficult: the same shifted family can yield lower localization error in some conditions while exposing severe failures in selected detectors. Generalization claims should therefore identify the physical and statistical changes represented by an evaluation condition rather than characterize it only as in-distribution or withheld.

Cross-grid transfer changes the physical domain itself. Under the evaluated protocols, \ac{fd} is comparatively portable without target-grid training, \ac{fc} benefits more consistently from target-grid adaptation, and \ac{fl} remains strongly dependent on target-grid examples. These differences show that transferability depends on the protection function; performance on one task cannot be taken as evidence that other protection capabilities will transfer similarly across grids.

Fine-tuning substantially improves over zero-shot transfer, but under matched preprocessing it shows no consistent final-performance advantage over target-only training. Pre-training therefore enables transfer before target-grid training without consistently improving the final result once target-grid data are available. The control also shows that preprocessing must be matched when attributing gains to pre-training. Transfer studies should distinguish zero-shot performance, gains from fine-tuning, and final target-grid performance.

\subsection*{Benchmark use and extensions}

EvEMTBench is shared evaluation infrastructure rather than a closed model competition. Comparability requires the benchmark artifacts associated with the reported evaluation cells to remain fixed, including task definitions, valid-sample masks, window construction, observability views, partitions, leakage controls, and metrics. New models can be evaluated under the existing benchmark specification, provided that training and model selection use only the prescribed partitions. Because the current release contains protocol-specific artifact versions, comparisons should report the version identifiers associated with the relevant protocol artifacts and result cells rather than assume that all results share a single version.

The benchmark can be extended without invalidating existing evaluation cells. Particularly relevant directions include sensor non-idealities, missing or delayed measurements, topology-held-out evaluation, and synthetic-to-real transfer. Such extensions would broaden the generalization questions addressed by EvEMTBench while preserving the existing reference evaluations. For competitive evaluations or repeated benchmark development, hidden test data may eventually be needed to preserve the independence of the final evaluation set.

\subsection*{Limitations and future directions}

The present evidence is based on synthetic \ac{emt} simulations designed for controlled comparison. It therefore does not establish performance on field measurements, deployment readiness, or operational event frequencies. Real protection systems introduce additional variability from equipment behavior, measurement chains, communication systems, and operating conditions that is not fully represented by the current simulations. Event coverage also differs across grids, so class-specific results should be interpreted within the corresponding grid and data family.

The current observability and model formulations introduce further limitations. Cross-grid transfer is restricted to local and line observability because the global input dimension varies with topology. In addition, the reference models receive waveform measurements without explicit line parameters, operating-point variables, network-strength indicators, or topology metadata. These choices provide controlled reference conditions but may limit localization and cross-grid transfer. Future evaluations should therefore include fixed-relay settings and models that use topology and grid parameters explicitly.

The baseline suite provides reproducible reference points rather than a performance ceiling. Established traveling-wave localization methods~\cite{abd_el-ghany_robust_2026} and stronger task-specific models would provide useful additional references. Finally, the observed feature-based failures and source-grid effects remain descriptive. For the \ac{rf}, the shifted-family failures may reflect changes in the magnitude-based summary features induced by the altered operating and event conditions, but the present evaluation does not isolate this mechanism. The cross-grid source comparisons likewise confound voltage level, system frequency, and source-corpus size. The matched-normalization control resolves the preprocessing confound in the fine-tuning comparison but does not explain why per-window normalization affects classification and localization. Dedicated single-factor experiments are required for the remaining source-grid patterns.

\section*{Conclusion}\label{sec:conclusion}

EvEMTBench establishes an open and reproducible benchmark for evaluating machine-learning-based power system protection across grids and generalization conditions. The reference evaluation shows that strong performance in one task or condition does not translate uniformly to others: the value of additional measurements depends on the protection function, shifted conditions can expose failures not apparent in-distribution, and cross-grid transfer is substantially stronger for fault detection than for fault localization. These findings show why protection models should be evaluated across clearly defined conditions rather than by isolated benchmark scores. By making such evaluations reproducible and comparable, EvEMTBench provides a shared reference point for future methods and allows evidence about generalization and failure modes to accumulate across studies. EvEMTBench is openly available to the research community, with the benchmark implementation, reference baselines, and evaluation resources released to support comparison and further development. In this way, EvEMTBench turns generalization from an implicit assumption into an explicit and repeatable evaluation problem for machine-learning-based power system protection.


\section*{Data availability}

The EvEMTBench Dataset used in this study is publicly available through FAUDataCloud at \url{https://data.fau.de/share/0e8d60feb7e65616c60aab78b93db77053275da53fd894bf5b75fc5e9ee7dfbf/}. The dataset is described by Kordowich et al., ``A Simulation Based Dataset of Faults and Events for Machine Learning in Power Systems'' (arXiv:2608.19777).

\section*{Code availability}

The EvEMTBench Benchmark implementation, committed reference splits, and reference results for the current release are publicly available under the MIT License at \url{https://github.com/EvEMTBench/evemtbench-benchmark}.

\section*{Use of generative AI and AI-assisted technologies}

During the preparation of this manuscript, the authors used ChatGPT (OpenAI) and Claude (Anthropic) to assist with drafting and language refinement of selected passages. All AI-assisted content was critically reviewed, verified, and edited by the authors. The authors remain fully responsible for the scientific reasoning, analyses, interpretation, conclusions, and final content of the manuscript.


\bibliography{references}


\section*{Supplementary information}

Supplementary Information provides complete held-out and cross-grid results, matched-control analyses, diagnostic results, dataset and grid details, baseline and software configurations, class support, and partition and validation-selection audits.


\section*{Acknowledgements}

This project was funded by the Deutsche Forschungsgemeinschaft (DFG, German Research Foundation) - 535389056.

\vspace{0.4cm}
\noindent The authors gratefully acknowledge the scientific support and high-performance computing resources provided by the Erlangen National High Performance Computing Center (NHR@FAU) at Friedrich-Alexander-Universit\"at Erlangen-N\"urnberg (FAU) under NHR project b304dc. NHR funding is provided by the German federal and Bavarian state authorities.


\section*{Author contributions statement}

J.O. conceived the benchmark, developed the methodology and software, conducted the validation, formal analysis, and investigation, curated the data, and generated the visualizations. G.K. contributed to the conception and methodology, generated the underlying data, and provided resources. C.B., A.M., J.J., and S.B. supervised the study. A.M., J.J., and S.B. administered the project and acquired funding. J.O. drafted the manuscript. All authors contributed to substantive revision of the manuscript and reviewed and approved the submitted version.

\section*{Competing interests}

The authors declare no competing interests.

\end{document}